\documentclass[conference]{IEEEtran}
\usepackage{amssymb}
\usepackage{amstext}
\usepackage{graphics}
\usepackage{epic,eepic,psfrag}
\usepackage{amsmath}

\usepackage{latexsym}
\usepackage{amssymb}
\usepackage{graphicx}
\usepackage{enumerate}
\usepackage{epsfig}
\usepackage{epsfig}
\usepackage{subfigure}
\usepackage{epstopdf}
\usepackage{epstopdf}
\usepackage{epsfig}
\usepackage{amsmath}
\usepackage{amssymb}
\usepackage{graphicx}
\usepackage{psfrag}

\usepackage{dblfloatfix}

\newcommand{\bA}{{\bf A}}
\newcommand{\bB}{{\bf B}}
\newcommand{\bC}{{\bf C}}
\newcommand{\bD}{{\bf D}}

\newcommand{\bF}{{\bf F}}
\newcommand{\bG}{{\bf G}}
\newcommand{\bH}{{\bf H}}
\newcommand{\bI}{{\bf I}}

\newcommand{\bK}{{\bf K}}

\newcommand{\bP}{{\bf P}}
\newcommand{\bQ}{{\bf Q}}
\newcommand{\bR}{{\bf R}}

\newcommand{\bU}{{\bf U}}
\newcommand{\bV}{{\bf V}}

\newcommand{\bX}{{\bf X}}
\newcommand{\bY}{{\bf Y}}

\newcommand{\bd}{{\bf d}}

\newcommand{\bq}{{\bf q}}
\newcommand{\br}{{\bf r}}
\newcommand{\bs}{{\bf s}}

\newcommand{\bw}{{\bf w}}
\newcommand{\bx}{{\bf x}}
\newcommand{\by}{{\bf y}}

\newcommand{\cC}{\mbox{$\cal C$}}
\newcommand{\cN}{\mbox{$\cal N$}}
\newcommand{\cH}{\mbox{$\cal H$}}
\DeclareMathOperator{\E}{\mathbb{E}}
\newcommand{\Hermitian}{{\rm H}}

\usepackage{bm}

\begin{document}

\title{Analysis of Discrete-Time MIMO OFDM-Based Orthogonal Time Frequency Space Modulation}

\author{\normalsize Ahmad RezazadehReyhani$^*$, Arman Farhang$^\dagger$, Mingyue Ji$^*$, Rong Rong Chen$^*$ and Behrouz Farhang-Boroujeny$^*$
	\\$^*$ECE Department, University of Utah, USA,
	$^\dagger$Trinity College Dublin, Ireland.\\
	Email: \{ahmad.rezazadeh, mingyue.ji, rchen, farhang\}@utah.edu, farhanga@tcd.ie}
\maketitle

\begin{abstract}
	Orthogonal Time Frequency Space (OTFS) is a novel modulation scheme designed in the Doppler-delay domain to fully exploit time and frequency diversity of general time-varying channels. In this paper, we present a novel discrete-time analysis of OFDM-based OTFS transceiver with a concise and vectorized input-output relationship that clearly characterizes the contribution of each underlying signal processing block in such systems. When adopting cyclic prefix in the time domain, our analysis reveals that the proposed MIMO OTFS and OFDM systems have the same ergodic capacity despite the well-known fact that the former has great advantages in low-complexity receiver design for high Doppler channels. The proposed discrete-time vectorized formulation is applicable to general fast fading channels with arbitrary window functions. It also enables practical low-complexity receiver design for which such a concise formulation of the input-output relationship is of great benefits.
\end{abstract}


\section{Introduction}\label{sec:Intro}
Orthogonal time frequency space (OTFS) modulation has recently been proposed as an effective waveform that takes advantage of the time diversity (i.e., variation of channel with time) to improve on the reliability of wireless links \cite{Hadani2017OTFS}. In OTFS, the transmit data symbols are treated as values of the grid points in a Doppler-delay space. A transformation step takes each data symbol and spread it over the entire space of time-frequency points. The result of this transformation is then passed to a multicarrier system for modulation and transmission. In this way, all data symbols are equally affected by the channel frequency selectivity and time diversity and, as a result, the time-varying channel, within a good approximation, converts to a unified time-invariant impulse response for all the data symbols.

To recover the transmitted (Doppler-delay space) data symbols, at the receiver, the respective multicarrier demodulation followed by a transformation takes the received signal back to the Doppler-delay space. In \cite{Hadani2017OTFS}, it is proposed that an inverse symplectic finite Fourier transform (SFFT$^{-1}$) be used for transformation from the Doppler-delay space to the time-frequency space, at the transmitter, and the corresponding symplectic finite Fourier transform (SFFT) be used for the reverse operation, at the receiver.

In the above set-up, the equivalent channel that connects the transmitted data symbols and the received signal samples, both in the Doppler-delay space, is modeled by a time-invariant two dimensional (2-D) impulse response. Soft detectors/equalizers, e.g., some extensions to those presented in \cite{Tuchler2002Turbo}, 
may thus be used for near-optimal recovery of the transmitted information. In order to simplify such detectors, \cite{Hadani2017OTFS} has proposed that proper windows should be applied to the time-frequency signals at both the transmitter and receiver sides to improve on the sparsity of the OTFS channel.

The transformation of data symbols from the Doppler-delay space to time-frequency at the transmitter and the corresponding inverse transformation at the receiver, clearly, leads to full diversity gain across both time and frequency. In multiple-input multiple-output (MIMO) channels, the space diversity gain will naturally be present because in a MIMO setup any signal going out of each antenna reaches all the receiver antennas with statistically similar gains.

The goal of this paper is to examine the details of OTFS in terms of the reliability of transmission brought up as a result of the addition of time diversity in the modulator. To allow simple derivations, OFDM is used for multicarrier transmission of time-frequency signals in OTFS. To this end, we present a novel discrete-time end-to-end formulation of an OFDM-based OTFS setup. Such a formulation provides a concise representation of the effect of each signal processing unit on the input-output relationship of an MIMO OFDM-based OTFS system. Based on this formulation, we show that the proposed MIMO OFDM-based OTFS system achieves the same ergodic capacity as that of an OFDM system, under the assumption that channel state information is known at the receiver and most importantly, with the use of cyclic prefix, a block of OTFS transmission can be implemented as $N$ consecutive, non-interfering OFDM transmissions. Since this capacity analysis is derived assuming perfect channel knowledge,  optimal receiver, and an infinite code length,  it does not violate the known fact that in practical systems with higher Dopplers, OTFS outperforms OFDM and has lower receiver complexity  due to simpler channel estimation and equalization  in the Doppler-delay domain.


This work is inspired by the landmark paper \cite{Hadani2017OTFS}, which was the first to propose a continuous-time formulation of a single-antenna OTFS modulation. Recently,  \cite{raviteja2017low}  presents a discrete-time formulation of a single-antenna OTFS system by sampling the continuous-time channel for the case when the transmit and receive window functions are rectangular. In \cite{li2017simple},  a matrix-form discrete-time formulation of a single-antenna transceiver with rectangular windows is presented. However, this formulation is mostly  based on two-dimensional signal matrices instead of vectors. Moreover, \cite{farhang2017low} studies a single-antenna transceiver when the transmit and receive window functions are separable. In comparison, our formulation here uses a general form of window functions in an MIMO OFDM-based OTFS setup and presents a vectorized formulation of signals which is amicable to analytical analysis and practical implementation of MIMO OTFS systems.


The rest of this paper is organized as follows. In Section \ref{sec:Kronecker}, we summarize basic properties of the Kronecker products for which
most of the derivations in this paper are based upon. In Sections ~\ref{sec:III} and ~\ref{sec:IV}, we describe the fundamentals of OTFS and present a vectorized formulation for the input-output relationship with general windowing functions for the single antenna case.  Extensions of the derivations to the MIMO case are presented in Section~\ref{sec:V}. The ergodic capacity analysis of the OFDM-based OTFS is developed in Section~\ref{sec:VI}. The paper is concluded in Section~\ref{sec:X}.

{\em Notations}:
Matrices, vectors, and scalar quantities are denoted by boldface uppercase, boldface lowercase, and normal letters, respectively. $\bI_N$ and $\b0_N$ are the identity and null matrices of size $N$, respectively, and $j$ is used to denote $\sqrt{-1}$. The superscripts $(\cdot)^{\rm T}$, $(\cdot)^{\rm H}$, and $(\cdot)^{\rm *}$ indicate transpose, conjugate transpose, and conjugate operations, respectively. The $N$-point discrete Fourier transform (DFT) matrix is represented by $\bF_N$ and is assumed to be normalized such that $\bF_N\bF_N^{\rm H}=\bI_N$. $\bA \in \cC^{M\times N}$ denotes a $M\times N$ matrix with complex-valued elements and $a_{m,n}$ represents element of matrix $\bA$ at $m$-th row and $n$-th column.

\section{Kronecker Product Properties}\label{sec:Kronecker}
In this paper, we make frequent use of the {\em Kronecker} product to arrive at a concise and elegant representation of the input-output relationship in OTFS systems. Hence, in this section we first review some of the relevant {\em Kronecker} properties. For an $m\times n$ matrix $\bA$ and a $p\times q$ matrix $\bB$, the {\em Kronecker} product $\bA \otimes \bB$ is a $mp\times nq$ block matrix
\begin{align}\label{eqn:kron-1}
	\bA \otimes \bB =
	\begin{bmatrix}
		a_{11}B & \cdots & a_{1n}B \\
		\vdots& \ddots & \vdots \\
		a_{m1}B & \cdots & a_{mn}B
	\end{bmatrix}.
\end{align}
The {\em Kronecker} product is an associative but non-commutative operator
\begin{align}\label{eqn:kron-2}
	\bA \otimes \bB &\neq \bB \otimes \bA, \\
	\left(\bA \otimes \bB\right) \otimes \bC &= \bA \otimes \left(\bB \otimes \bC\right).
\end{align}
The mixed-product property of the {\em Kronecker} product is
\begin{align}\label{eqn:kron-3}
	\left(\bA \otimes \bB \right)\left(\bC \otimes \bD \right) = \left(\bA\bC \right) \otimes \left(\bB\bD \right).
\end{align}
Moreover, under transpose and Hermitian operations,  order of the {\em Kronecker} product operands does not change
\begin{align}\label{eqn:kron-4}
\left(\bA \otimes \bB \right)^{\rm H} = \bA^{\rm H} \otimes \bB^{\rm H}.
\end{align}
Finally, consider a matrix equation $\bA \bX \bB = \bC$, where $\bA$, $\bX$, $\bB$, and $\bC$ are proper size matrices. We can rewrite this equation as
\begin{align}\label{eqn:kron-5}
	\left( \bB^{\rm T} \otimes \bA \right) \operatorname {vec}(\bX) = \operatorname {vec}(\bC),
\end{align}
where $\operatorname {vec}(\bX)$ denotes the vectorized version of the matrix $\bX$ formed by stacking the columns of $\bX$ into a single column vector.

\begin{figure*}
	\psfrag{x}{\hspace{-1 mm}\small $d_{m,n}$}
	\psfrag{y}{\hspace{2 mm}\small $\widetilde{x}_{k,l}$}
	\psfrag{z}{\hspace{2 mm}\small $y_{k,l}$}
	\psfrag{ +}{\hspace{6.5 mm}\scriptsize $+$}
	\psfrag{xhat}{\small $\hat{d}_{m,n}$}
	\psfrag{Ch}{\small LTV Channel}
	\psfrag{Window}{\hspace{3.5 mm}\small SFFT}
	\psfrag{SFFTinv}{\hspace{1 mm} \small SFFT$^{-1}$}
	\psfrag{Tx}{\hspace{2 mm}\small ~OFDM-based OTFS Transmitter}
	\psfrag{Rx}{\hspace{3.5 mm}\small ~OFDM-based OTFS Receiver}
	\psfrag{OFDM_Tx}{\small\hspace{-0.1cm}OFDM}
	\psfrag{OFDM}{\hspace{-3 mm}\small OFDM}
	\psfrag{Demod}{\hspace{-6 mm}\small Demodulator}
	\psfrag{Mod}{\hspace{-5.5 mm}\small Modulator}
	\psfrag{SFFT}{\hspace{0.2 mm}\small Windowing }
	\psfrag{tWin}{\hspace{0.0 mm}\small Windowing }
	\centering
	\includegraphics[scale=0.28]{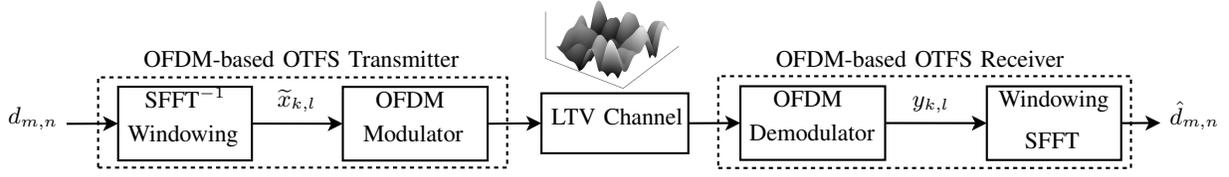}
	\caption{OTFS transmitter and receiver structure.}
	\vspace{-2mm}
	\label{fig:OTFS}
\end{figure*}
\section{OTFS Waveform}\label{sec:III}
An OTFS transmitter combines a set of complex-valued data symbols $\{d_{m,n}; m=0,1,\cdots,M-1; n=0,1,\cdots N-1\}$ in the Doppler-delay domain to construct an OTFS signal. First, the transmitter maps  data symbols on the Doppler-delay lattice, $\{d_{m,n}\}$, to a lattice in the time-frequency domain through an inverse symplectic finite Fourier transform operation (SFFT$^{-1}$)
\begin{align}\label{eqn:OTFS-1}
	x_{k,l} =\frac{1}{\sqrt{MN}} \sum_{m=0}^{M-1}\sum_{n=0}^{N-1} d_{m,n} e^{-j2\pi(\frac{mk}{M}-\frac{nl}{N})},
\end{align}
where $k=0,1,\cdots,M-1$ and $l=0,1,\cdots N-1$. Let $\bD \in \cC^{M\times N}$ denote the data matrix    which contains elements  $\{d_{m,n}\}$. A closer look into (\ref{eqn:OTFS-1}) reveals that the SFFT$^{-1}$ transform of  $\bD$ can be obtained by applying an $M$-point discrete Fourier transform (DFT) and an $N$-point inverse DFT (IDFT) to the columns and rows of the matrix $\bD$, respectively. Accordingly, (\ref{eqn:OTFS-1}) can be  written in a compact form as
\begin{align}\label{eqn:OTFS-2}
\bX = \bF_M \bD \bF_N^{\rm H},
\end{align}
where $\bX \in \cC^{M\times N}$.

Subsequently, the transmitter applies a transmit window, $\widetilde{u}_{k,l}$, to the time-frequency signal as
\begin{align}\label{eqn:OTFS-3}
	\widetilde{x}_{k,l} = x_{k,l} \widetilde{u}_{k,l}.
\end{align}

Finally, the transmitter packs the time-frequency signal, $\widetilde{x}_{k,l}$, to the transmitted signal, $s[i]$, using a set of time-frequency basis functions, $g_{k,l}[i]$,
\begin{align}\label{eqn:OTFS-4}
	s[i] = \sum_{k=0}^{M-1}\sum_{l=0}^{N-1}\widetilde{x}_{k,l}g_{k,l}[i].
\end{align}
The received signal samples after transmission over a linear time varying (LTV) channel can be obtained as
\begin{align}\label{eqn:OTFS-5}
	r[i] = \sum_{l=0}^{L-1}h[i,l]s[i-l] + w[i],
\end{align}
where $h[i,l]$ is the instantaneous channel impulse response with length $L$ at time instant $l$, and $w[i]$ is the additive channel noise.

To obtain an estimate of $d_{m,n}$ from  the received signal $r[i]$, first, we map  $r[i]$ to the time-frequency lattice by projecting it onto another set of time-frequency basis functions $f_{k,l}[i]$ as
\begin{align}\label{eqn:OTFS-6}
	\widetilde{y}_{k,l} = <f_{k,l}[i],r[i]>.
\end{align}	
The sets of transmit and receive  basis functions are designed to satisfy the perfect reconstruction criterion
\begin{align}\label{eqn:OTFS-7}
	<f_{k,l}[i],g_{k',l'}[i]> = \delta(k-k')\delta(l-l').
\end{align}
Then, the receiver performs the windowing operation through multiplication of the receive window function coefficients $\widetilde{v}_{k,l}$ to the time-frequency samples as
\begin{align}\label{eqn:OTFS-8}
	y_{k,l} = \widetilde{y}_{k,l} \widetilde{v}_{k,l}.
\end{align}

Finally,  signals on the time-frequency lattice are transformed back to the Doppler-delay domain by applying SFFT as
\begin{align}\label{eqn:OTFS-9}
	\hat{d}_{m,n} =\frac{1}{\sqrt{MN}} \sum_{k=0}^{M-1}\sum_{l=0}^{N-1} y_{k,l} e^{j2\pi(\frac{mk}{M}-\frac{nl}{N})}.
\end{align}
Similar to SFFT$^{-1}$, the SFFT transform can be split into an $M$-point IDFT and an $N$-point DFT on rows and columns of its operand. Thus, (\ref{eqn:OTFS-9}) can be rearranged as
\begin{align}\label{eqn:OTFS-10}
\hat{\bD} = \bF_M^{\rm H} \bY \bF_N,
\end{align}
where $\bY \in \cC^{M\times N}$ and $\hat{\bD}  \in \cC^{M\times N}$ contain elements $\{y_{k,l}\}$ and $\{\hat{d}_{m,n}\}$, respectively.

Typically, an OFDM-based OTFS modulation with $M$ subcarriers and a cyclic prefix (CP) of length $M_{\rm cp}$ utilizes the following transmit and receive time-frequency basis functions
\begin{align}\label{eqn:OTFS-11}
g_{k,0}[i] &= \frac{1}{\sqrt{M}}e^{j\frac{2\pi}{M}ki}, ~~ -M_{\rm cp}\leq i< M \nonumber \\
g_{k,l}[i] &= g_{k,0}[i-\left(M+M_{\rm cp}\right)l] \\
f_{k,0}[i] &= \frac{1}{\sqrt{M}}e^{-j\frac{2\pi}{M}ki},~~ 0\leq i< M \nonumber \\
f_{k,l}[i] &= f_{k,0}[i-\left(M+M_{\rm cp}\right)l].
\end{align}

\section{OFDM-Based OTFS: Single Antenna}\label{sec:IV}
In this section, we present a vectorized formulation for the OFDM-based OTFS transceiver.
The first step of the OTFS modulation in (\ref{eqn:OTFS-2}), i.e. the SFFT$^{-1}$ operation, using (\ref{eqn:kron-5}) can be rearranged as
\begin{align}\label{eqn:SISO-1}
 	\bx = \left(\bF_N^{\Hermitian} \otimes \bF_M \right) \bd,
\end{align}
where $\bx = \operatorname {vec}(\bX)$ and $\bd = \operatorname {vec}(\bD)$.
Then, the vector $\bx$ is partitioned into blocks of length $M$, denoted by $\bx_n,~n=0,1,\cdots,N-1$ and each block is multiplied to the corresponding transmit window as
\begin{align}\label{eqn:SISO-2}
	\widetilde{\bx}_n = \bU_n \bx_n,
\end{align}
where the transmit window $ \bU_n \in \cC^{M\times M}$ is a diagonal matrix consisting of diagonal elements $\widetilde{u}_{k,n}$.
 Stacking the results as $\widetilde{\bx}$, we have
\begin{align}\label{eqn:SISO-3}
	\widetilde{\bx} &= \bU \bx 
\end{align}
where $\bU \in \cC^{MN\times MN}$ is a diagonal matrix whose $(lM+k)$-th diagonal element is $\widetilde{u}_{k,l}$.

Each partition of $\widetilde{\bx}$, i.e. $\widetilde{\bx}_n$ , is fed into an OFDM modulator. The OFDM modulator multiplies an IDFT matrix $\bF_M^{\rm H}$ to each block and then appends a CP to each block as
\begin{align}\label{eqn:SISO-4}
	\bs_n &=  \bF_M^{\rm H} \widetilde{\bx}_n, \\
	\widetilde{\bs}_n &= \bA_{\rm cp} \bs_n \label{eqn:OFDM-sb},
\end{align}
where $M_{\rm cp}$ is the CP length, $\bA_{\rm cp} = [\bG_{\rm cp},\bI_M]^{\rm T}$ is the CP addition matrix and $\bG_{\rm cp}  \in \cC^{M\times M_{\rm cp}}$ includes the last $M_{\rm cp}$ columns of the identity matrix $\bI_M$. Note that $\bA_{\rm cp}$ appends last $M_{\rm cp}$ samples of each OFDM block to its beginning. Stacking the results as vectors, (\ref{eqn:SISO-4}) and (\ref{eqn:OFDM-sb}) can be written as
\begin{align}\label{eqn:SISO-5}
\bs &=  \left(\bI_N \otimes \bF_M^{\rm H} \right) \widetilde{\bx},\\
\widetilde{\bs} &=  \left(\bI_N \otimes \bA_{\rm cp} \right) \bs.\label{eqn:OTFS-sb}
\end{align}

After the signal $\widetilde{\bs}$ is passed though the LTV channel, the received signal can be written as
\begin{align}\label{eqn:SISO-6}
\widetilde{\br} = \bH \widetilde{\bs} + \bw,
\end{align}
where $\bH \in \cC^{N(M+M_{\rm cp}) \times N(M+M_{\rm cp})}$ is the channel impulse response matrix, and $\bw$ is the channel additive noise vector.
The receiver partitions the received vector into $N$ blocks, $\widetilde{\br}_n$ and removes CP from each received block as
\begin{align}\label{eqn:SISO-7}
	\br_n = \bR_{\rm cp} \widetilde{\br}_n,
\end{align}
where $\bR_{\rm cp} \in \cC^{M \times (M+M_{\rm cp})}$ is the CP removal matrix which can be obtained by removing the first $M_{\rm cp}$ rows of $\bI_{M+M_{\rm cp}}$.

Stacking the output vectors $\br_n$ into a length $NM$ vector $\br$, (\ref{eqn:OTFS-sb}) to (\ref{eqn:SISO-7}) can be written as
\begin{align}\label{eqn:SISO-8}
	\br &=  \left(\bI_N \otimes \bR_{\rm cp}\right) \bH \left(\bI_N \otimes \bA_{\rm cp}\right)\bs + \bw \nonumber\\
		&= \widetilde{\bH} \bs +\bw,
\end{align}
where $\widetilde{\bH}=\left(\bI_N \otimes \bR_{\rm cp}\right) \bH \left(\bI_N \otimes \bA_{\rm cp}\right)$ is an $MN\times MN$ block diagonal matrix
\begin{align}\label{eqn:SISO-9}
	\widetilde{\bH} &= \begin{bmatrix}
		\widetilde{\bH}_0 &  \b0_M      &  \cdots& \b0_M  		\\
		\b0_M       & \widetilde{\bH}_1 & \cdots & \b0_M  		\\
		\vdots      & \vdots      & \ddots & \vdots 		\\
		\b0_M       & \b0_M       & \cdots & \widetilde{\bH}_{N-1}
	\end{bmatrix}
\end{align}
and $\widetilde{\bH}_n$ is the channel impulse response matrix of the $n$-th OFDM symbol.

The received signal is fed into an OFDM demodulator and the output signal is
\begin{align}\label{eqn:SISO-10}
	\widetilde{\by} =  \left(\bI_N \otimes \bF_M \right)  \br,
\end{align}
where $\widetilde{\by}$ is a length $MN$ time-frequency signal vector. Finally, after performing the receiver windowing operation, the time-frequency signal is mapped to the Doppler-delay domain and an estimate of the transmitted vector can be obtained as
\begin{align}\label{eqn:SISO-11}
\hat{\bd} = \left(\bF_N \otimes \bF_M^{\Hermitian} \right) \by,
\end{align}
where $\by = \bV \widetilde{\by}$, and the receive window $\bV\in \cC^{MN\times MN}$ is a diagonal matrix defined similarly to that of the transmit window.

 We note that equations
(\ref{eqn:SISO-1}) to (\ref{eqn:SISO-11}) describe the end-to-end relationship for a general OFDM-based OTFS setup which is summarized in a compact form shown in  (\ref{eqn:SISO-11.1}), at the top of the next page. It clearly demonstrates the contribution of each  signal processing step of an OFDM-based OTFS setup from Doppler-delay domain to Doppler-delay domain where each step corresponds to one section of the transceiver block diagram shown in Fig. \ref{fig:OTFS}. It is clear that this compact form of (\ref{eqn:SISO-11.1}) will allow simplified analysis and implementation of OFDM-based OTFS systems.
Next, we will present how these equations simplify for some special setup considered below.
\begin{figure*}[!h]	
	\begin{align}\label{eqn:SISO-11.1}
	\hat{\bd} = \underbrace{\left(\bF_N \otimes \bF_M^{\Hermitian} \right)}_\text{SFFT}\overbrace{\bV}^\text{window} \underbrace{\left(\bI_N \otimes \bF_M \right)}_\text{OFDM Demod.}\overbrace{\left(\bI_N \otimes \bR_{\rm cp}\right)}^\text{Remove CP}\underbrace{ \bH}_\text{Channel} \overbrace{\left(\bI_N \otimes \bA_{\rm cp}\right)}^\text{Add CP} \underbrace{\left(\bI_N \otimes \bF_M^{\rm H} \right)}_\text{OFDM Mod.} \overbrace{\bU}^\text{Window} \underbrace{\left(\bF_N^{\Hermitian} \otimes \bF_M \right)}_\text{Inv. SFFT} \bd+\hat{\bw}.
	\end{align}
	\hrulefill	
\end{figure*}
\subsection{Separable Windows}
In general, transmit and receive window functions can be arbitrary functions
that are designed for various purposes. These include, for instance,   zero-forcing or minimum mean squared error equalization, or shortening the channel response in Doppler-delay domain.  For a wide range of applications, however, one can treat windowing across time axis and across frequency axis separately because of the independency between channel delay spread and Doppler spread.  This separate windowing can lead to simplified transmitter and receiver structures  as shown in \cite{farhang2017low}.

 Assume that the transmit and receive windows are separable functions, i.e. $\widetilde{u}_{k,l} = a_lb_k$ and $\widetilde{v}_{k,l} = p_lq_k$. We can write transmit and receive window matrices as $\bU = \bA\otimes\bB$ and $\bV = \bP\otimes\bQ$ where $\bA \in \cC^{N\times N}$, $\bP \in \cC^{N\times N}$, $\bB \in \cC^{M\times M}$, and $\bQ \in \cC^{M\times M}$  are diagonal matrices with elements $\{a_l\}$, $\{p_l\}$,  $\{b_k\}$ ,and $\{q_k\}$, respectively. Substituting these window forms in (\ref{eqn:SISO-3}) to (\ref{eqn:SISO-11}) and after simplifications we have
\begin{align}\label{eqn:SISO-12}
\hat{\bd} =  &\left(\bI_N \otimes \bF^{\rm H}_M \bQ \bF_M\right) \left(\bF_N \otimes \bI_M\right) \left(\bP \otimes \bI_M\right) \widetilde{\bH} \nonumber \\ &\left( \bA \otimes \bI_M \right)\left(\bF^{\rm H}_N \otimes \bI_M \right)\left( \bI_N \otimes \bF^{\rm H}_M \bB \bF_M \right) \bd +\hat{\bw},
\end{align}
where $\hat{\bw} = \left(\bF_N \otimes \bF_M^{\Hermitian} \right) \bV \left(\bI_N \otimes \bF_M \right) \bw$. We note that (\ref{eqn:SISO-12}) can be used for the simplification of OFDM-based OTFS transceiver.

\subsection{Rectangular Window}
When rectangular windows are used as the transmitter and receiver window functions, i.e. $\bU = \bV = \bI_{MN}$, substituting (\ref{eqn:SISO-3}) to (\ref{eqn:SISO-10}) in (\ref{eqn:SISO-11}), it simplifies to
\begin{align}\label{eqn:SISO-13}
\hat{\bd} &= \left(\bF_N \otimes \bI_M \right) \widetilde{\bH} \left(\bF^{\rm H}_N \otimes \bI_M \right) \bd + \hat{\bw},
\end{align}
where we used the mixed-product property of Kronecker product $\left(\bA\otimes \bB\right)\left(\bC\otimes \bD\right) = \left(\bA\bC\right)\otimes \left(\bB\bD\right)$.

\subsection{Frequency Domain Representation}
Let $\widetilde{\bH}_f$  denote the block diagonal frequency domain channel impulse response matrix. We see that $\widetilde{\bH}_f = \left( \bI_N\otimes \bF_M \right) \widetilde{\bH} \left( \bI_N\otimes \bF^{\rm H}_M \right)$.  Solving this for $\widetilde{\bH}$ we obtain $\widetilde{\bH}=\left( \bI_N\otimes \bF^{\rm H}_M \right) \widetilde{\bH}_f \left( \bI_N\otimes \bF_M \right)$. Substituting $\widetilde{\bH}$ in (\ref{eqn:SISO-11.1}), after simplification it boils down to
\begin{align}\label{eqn:SISO-14}
\hat{\bd} &= \left(\bF_N \otimes \bF^{\rm H}_M \right) \bV  \widetilde{\bH}_f  \bU \left(\bF^{\rm H}_N \otimes \bF_M \right) \bd + \hat{\bw},
\end{align}
which shows that transmit and receive windows directly change the frequency domain channel impulse response matrix. In other words, the effective frequency domain channel is $\bV  \widetilde{\bH}_f  \bU$.

When the LTV channel  varies slowly  such that it is approximately invariant over each OFDM symbol, then the matrices $\{\widetilde{\bH}_n\}$ become circulant. Accordingly, $\widetilde{\bH}_f$ is an $MN \times MN$ diagonal matrix that contains $N$ diagonal channel frequency response matrices, $\widetilde{\bH}_{f,n} = \bF_M^{\rm H} \widetilde{\bH}_n \bF_M$. In this case, (\ref{eqn:SISO-14}) is equivalent to the 2D circular convolution of the transmit data matrix with the effective 2D channel impulse response matrix.


\section{OFDM-Based OTFS: MIMO}\label{sec:V}
Consider a $n_{\rm t}\times n_{\rm r}$ MIMO setup that utilizes OFDM-based OTFS modulation. Let us stack transmit data matrices of all antennas to form an $Mn_{\rm t}\times N$ matrix $\overline{\bD}$ as
\begin{align}\label{eqn:MIMO-1}
\overline{\bD}=\left[\hspace{-2mm} \begin{array}{c}
\bD^0\\
\bD^1\\
\vdots\\
\bD^{n_{\rm t}-1}\end{array} \hspace{-2mm} \right],
\end{align}
where $\bD^t$ is the $M\times N$ data matrix of $t$-th antenna. Applying SFFT$^{-1}$ to each sub-matrix $\bD^t$, we obtain
\begin{align}\label{eqn:MIMO-2}
\overline{\bX}= \left(\bI_{n_{\rm t}} \otimes \bF_M \right) \overline{\bD} \bF_N^{\Hermitian},
\end{align}
where $\bF_M$ is repeated as $\bI_{n_{\rm t}} \otimes \bF_M$ to cover all data matrices of all antennas. Using (\ref{eqn:kron-5}), (\ref{eqn:MIMO-2}) can be written as
\begin{align}\label{eqn:MIMO-3}
\overline{\bx} = \left(\bF_N^{\Hermitian} \otimes \bI_{n_{\rm t}} \otimes \bF_M \right) \overline{\bd},
\end{align}
where $\overline{\bd}$ is the vectorized version of $\overline{\bD}$ defined as follows.
\begin{align}\label{eqn:MIMO-4}
\overline{\bd}=\left[\hspace{-2mm} \begin{array}{c}
\bd_0\\
\bd_1\\
\vdots\\
\bd_{N-1}\end{array} \hspace{-2mm} \right],
~
\bd_n=\left[ \hspace{-2mm} \begin{array}{c}
\bd_{n}^0\\
\bd_{n}^1\\
\vdots\\
\bd_{n}^{{n_{\rm t}}-1}\end{array} \hspace{-2mm} \right],
~
\bd_n^t=\left[ \hspace{-2mm} \begin{array}{c}
d_{0,n}^t\\
d_{1,n}^t\\
\vdots\\
d_{M-1,n}^t\end{array} \hspace{-2mm} \right],
\end{align}
and $d_{m,n}^t; t=0,1,\cdots,{n_{\rm t}}-1$ is the data symbol of $t$-th transmit antenna.

After partitioning $\overline{\bx}$ to $N$ blocks, $\overline{\bx}_n$, the window function is multiplied to each block. Assuming that all transmit antennas use the same window function, we get
\begin{align}\label{eqn:MIMO-5}
\breve{\bx}_n &= \overline{\bU}_n \overline{\bx}_n,
\end{align}
where $\overline{\bU}_n = \bI_{n_{\rm t}} \otimes \bU_n$. We  stack these results to obtain
\begin{align}\label{eqn:MIMO-6}
\breve{\bx}   &= \overline{\bU} \overline{\bx},
\end{align}
where
\begin{align}\label{eqn:MIMO-7}
\overline{\bU} &= \begin{bmatrix}
\overline{\bU}_0 &  \b0_{M{n_{\rm t}}}           & \cdots & \b0_{M{n_{\rm t}}}  		\\
\b0_{M{n_{\rm t}}}            & \overline{\bU}_1 & \cdots & \b0_{M{n_{\rm t}}}  		\\
\vdots              & \vdots              & \ddots & \vdots 		\\
\b0_{M{n_{\rm t}}}            & \b0_{M{n_{\rm t}}}            & \cdots & \overline{\bU}_{N-1}
\end{bmatrix}.
\end{align}

The OFDM modulator transforms the time-frequency domain signal of each antenna to the time domain signal as
\begin{align}\label{eqn:MIMO-8}
\bs_n =  \left(\bI_{n_{\rm t}} \otimes \bF_M^{\rm H} \right) \breve{\bx}_n,\\
\bs =  \left(\bI_{N{n_{\rm t}}} \otimes \bF_M^{\rm H} \right) \breve{\bx}.
\end{align}
The transmitter appends the CP to the transmit signal and sends the result through an LTV channel. At the receiver, after removing the CP, the received signal can be written in a compact form as
\begin{align}\label{eqn:MIMO-9}
\br_n &=  \left(\bI_{n_{\rm r}} \otimes \bR_{\rm cp}\right) \bH_n \left(\bI_{n_{\rm t}} \otimes \bA_{\rm cp}\right)\bs_n + \bw_n \nonumber\\
&= \overline{\bH}_n \bs_n +\bw_n,
\end{align}
where $\overline{\bH}_n=\left(\bI_{n_{\rm r}} \otimes \bR_{\rm cp}\right) \bH_n \left(\bI_{n_{\rm t}} \otimes \bA_{\rm cp}\right)$ can be expanded as
\begin{align}\label{eqn:MIMO-10}
\overline{\bH}_n &= \begin{bmatrix}
\overline{\bH}_n^{0,0}       & \overline{\bH}_n^{0,1}   & \cdots & \overline{\bH}_n^{0,{n_{\rm t}}-1} \\
\overline{\bH}_n^{1,0}       & \overline{\bH}_n^{1,1}   & \cdots & \overline{\bH}_n^{1,{n_{\rm t}}-1} \\
\vdots                  & \vdots              & \ddots & \vdots 		\\
\overline{\bH}_n^{{n_{\rm r}}-1,0}     & \overline{\bH}_n^{{n_{\rm r}}-1,1} & \cdots & \overline{\bH}_n^{{n_{\rm r}}-1,{n_{\rm t}}-1}
\end{bmatrix}.
\end{align}
Here, $\overline{\bH}_n^{r,t}$ is the $n$-th channel impulse response matrix between $t$-th transmit and $r$-th receive antenna.

We stack the output vectors to obtain
\begin{align}\label{eqn:MIMO-11}
\br &=  \overline{\bH} \bs +\bw,
\end{align}
where
\begin{align}\label{eqn:MIMO-12}
	\overline{\bH} &= \begin{bmatrix}
	\overline{\bH}_0             & \b0_{M{n_{\rm r}}\times M{n_{\rm t}}}   & \cdots & \b0_{M{n_{\rm r}}\times M{n_{\rm t}}}  		\\
	\b0_{M{n_{\rm r}}\times M{n_{\rm t}}}       & \overline{\bH}_1         & \cdots & \b0_{M{n_{\rm r}}\times M{n_{\rm t}}}  		\\
	\vdots                  & \vdots              & \ddots & \vdots 		\\
	\b0_{M{n_{\rm r}}\times M{n_{\rm t}}}       & \b0_{M{n_{\rm r}}\times M{n_{\rm t}}}   & \cdots & \overline{\bH}_{N-1}
	\end{bmatrix}.
\end{align}

The received signal is fed to an OFDM demodulator and the output signal is
\begin{align}\label{eqn:MIMO-13}
\breve{\by} =  \left(\bI_{N{n_{\rm r}}} \otimes \bF_M \right)  \br,
\end{align}
where $\breve{\by}$ is the length $MN{n_{\rm r}}$ time-frequency signal vector. Finally, after multiplying by the receiver window, the time-frequency signal is mapped to the Doppler-delay domain and the estimate of the transmitted vector can be obtained as
\begin{align}\label{eqn:MIMO-14}
\hat{\bd} = \left(\bF_N \otimes \bI_{n_{\rm r}} \otimes \bF_M^{\Hermitian} \right) \overline{\bV} \breve{\by},
\end{align}
where
\begin{align}\label{eqn:MIMO-15}
	\overline{\bV} &= \begin{bmatrix}
	\overline{\bV}_0 &  \b0_{M{n_{\rm r}}}           & \cdots & \b0_{M{n_{\rm r}}}  		\\
	\b0_{M{n_{\rm r}}}            & \overline{\bV}_1 & \cdots & \b0_{M{n_{\rm r}}}  		\\
	\vdots              & \vdots              & \ddots & \vdots 		\\
	\b0_{M{n_{\rm r}}}            & \b0_{M{n_{\rm r}}}            & \cdots & \overline{\bV}_{N-1}
\end{bmatrix}
\end{align}
and $\overline{\bV}_n = \bI_{n_{\rm r}} \otimes \widetilde \bV_n$. Using (\ref{eqn:MIMO-3}) to (\ref{eqn:MIMO-14}), it is straightforward to find a similar  end-to-end vectorized relationship for the general MIMO OFDM-based OTFS setup to that of (\ref{eqn:SISO-11.1}).


\section{Capacity Analysis of \\MIMO OFDM-Based OTFS}\label{sec:VI}
In this section, we examine the ergodic capacity of MIMO-OFDM and MIMO-OFDM-based OTFS systems. While the analysis here applies to general time-varying channels, it is important to note that due to the use of cyclic prefix, the transmission of an OTFS block of symbols consists of $N$ consecutive transmissions of OFDM blocks, each with a block length $M+M_{\rm cp}$ in the time domain. Assuming that the receiver knows the channel perfectly, one (larger) block of OTFS transmission is equivalent to $N$ parallel transmissions of OFDM blocks. Furthermore, we assume that the channel is independent from one OTFS block to the next and the channel is ergodic. These form the underlining key assumptions of the capacity analysis below.


Consider an arbitrary OTFS transmission block. Since there is a one-to-one mapping between Doppler-delay domain data symbols $\overline{\bd}$ and time-frequency data samples $\overline{\bx}$, the mutual information between the OTFS received signal vector $\br$ and transmit data vector $\overline{\bd}$, i.e. $I(\overline{\bd}; \br)$, can be written as
\begin{align}\label{eqn:CAP-0}
	I(\overline{\bd};\br) = I(\overline{\bx}_0,\overline{\bx}_1,\cdots,\overline{\bx}_{N-1};\br_0,\br_1,\cdots,\br_{N-1}).
\end{align}
Since the received signal $\br_n$ only depends on $\overline{\bx}_n$, i.e. $\br_n= \cH_n \overline{\bx}_n+\bw_n$, where $\cH_n$ represents all signal processing steps that relates $\overline{\bx}_n$ to $\br_n$ including modulation and channel impact. Then, (\ref{eqn:CAP-0}) can be written as
\begin{align}\label{eqn:CAP-01}
I(\overline{\bd};\br) = \sum_{n=0}^{N-1} I(\overline{\bx}_n;\br_n),
\end{align}
 due to the fact that each OTFS transmission is realized by transmissions over a set of $N$ parallel channels. Furthermore, each parallel channel is an MIMO channel $\br_n= \cH_n \overline{\bx}_n + \bw_n $ with known channel state information at the receiver. Since the capacity-achieving input distribution for such a channel is the zero-mean circularly symmetric complex Gaussian distribution $\cC\cN(0,\bI_M)$, \cite{tse2005fundamentals}, we conclude that
such distribution is indeed capacity-achieving for the proposed MIMO-OFDM based OTFS system and the resulting ergodic capacity equals to the average sum capacities of individual parallel channels, which equals with the ergodic capacity of the OFDM system considered here \cite{tse2005fundamentals}.

In the following, for an OFDM-based OTFS setup we calculate $I(\overline{\bd};\br)$ and show that (\ref{eqn:CAP-01}) holds for OFDM-based OTFS modulation and we proceed to calculate the capacity of OFDM-based OTFS setup.
We recall that the differential entropy of an $n$-element complex Gaussian vector $\bq$ with covariance matrix $\bC_{\bq}$ is  \cite{cover2012elements}
\begin{equation}
\label{eqn:CAP-1}
h(\bq) = \log_2 \left((2\pi e)^n |\bC_{\bq}|\right)
\end{equation}
where $|\cdot|$ is the determinant operator. Moreover, the mutual information between two complex Gaussian vectors $\bq$ and $\br$ is the difference between the differential entropies $h(\bq)$ and $h(\bq|\br)$. This is written as
\begin{align}\label{eqn:CAP-2}
I(\bq;\br) 	&= h(\bq) -  h(\bq|\br) \nonumber\\
&= \log_2 \frac{|\bC_{\bq}|}{|\bC_{\bq|\br}|}.
\end{align}

Substituting (\ref{eqn:MIMO-5}) and (\ref{eqn:MIMO-8}) in (\ref{eqn:MIMO-9}), the received signal of the $n$-th OFDM transmission can be obtained as
\begin{align}\label{eqn:CAP-3}
	\br_n &= \overline{\bH}_n\left(\bI_{n_{\rm t}} \otimes \bF_M^{\rm H} \right) \overline{\bU}_n \overline{\bx}_n +\bw_n.
\end{align}
Then, the mutual information between the OFDM received signal vector $\br_n$ and the transmit data vector $\overline{\bx}_n$ is
\begin{align} \label{eqn:CAP-4}
I(\br_n;\overline{\bx}_n) = \log_2\left(\frac{|\bK_n \bI_{M{n_{\rm t}}} \bK_n^{\rm H} +\sigma^2\bI_{M{n_{\rm r}}}|}{|\sigma^2\bI_{M{n_{\rm r}}}|}\right)
\end{align}
where $\bK_n = \overline{\bH}_n\left(\bI_{n_{\rm t}} \otimes \bF_M^{\rm H} \right) \overline{\bU}_n$.

Similarly, the received signal of OFDM-based OTFS can be obtained by substituting (\ref{eqn:MIMO-3}) and (\ref{eqn:MIMO-6}) in (\ref{eqn:MIMO-11}) as
\begin{align}\label{eqn:CAP-5}
\br &= \overline{\bH}\left(\bI_{N{n_{\rm t}}} \otimes \bF_M^{\rm H} \right) \overline{\bU} \left(\bF_N^{\Hermitian} \otimes \bI_{n_{\rm t}} \otimes \bF_M \right) \overline{\bd} +\bw.
\end{align}

The mutual information between the OFDM-based OTFS received signal vector $\br$ and the transmit data vector $\overline{\bd}$ is given as
\begin{align}\label{eqn:CAP-7}
I(\br;\overline{\bd}) = \log_2\left(\frac{|\bK \bI_{MN{n_{\rm t}}} \bK^{\rm H} +\sigma^2\bI_{MN{n_{\rm r}}}|}{|\sigma^2\bI_{MN{n_{\rm r}}}|}\right)
\end{align}
where $\bK = \overline{\bH}\left(\bI_{N{n_{\rm t}}} \otimes \bF_M^{\rm H} \right) \overline{\bU} \left(\bF_N^{\Hermitian} \otimes \bI_{n_{\rm t}} \otimes \bF_M \right)$.

Note that $\bF_N^{\Hermitian} \otimes \bI_{n_{\rm t}} \otimes \bF_M$ is a unitary matrix and $\overline{\bH}\left(\bI_{N{n_{\rm t}}} \otimes \bF_M^{\rm H} \right) \overline{\bU}$ is a block diagonal matrix, thus we have
\begin{align}\label{eqn:CAP-8}
\bK \bK^{\rm H} &= \begin{bmatrix}
\bK_0 \bK_0^{\rm H}           & \cdots & \b0_{M{n_{\rm r}}}  		\\
\b0_{M{n_{\rm r}}}            & \bK_1 \bK_1^{\rm H} & \cdots & \b0_{M{n_{\rm r}}}  		\\
\vdots              & \vdots              & \ddots & \vdots 		\\
\b0_{M{n_{\rm r}}}            & \b0_{M{n_{\rm r}}}            & \cdots & \bK_{N-1}\bK_{N-1}^{\rm H}
\end{bmatrix}.
\end{align}
Substituting (\ref{eqn:CAP-8}) in (\ref{eqn:CAP-7}), we have

\begin{align}\label{eqn:CAP-9}
I(\br;\overline{\bd}) &= \sum_{n=0}^{N-1} \log_2\left(\frac{|\bK_n \bK_n^{\rm H} +\sigma^2\bI_{M{n_{\rm r}}}|}{|\sigma^2\bI_{M{n_{\rm r}}}|}\right) \nonumber \\
				 &= \sum_{n=0}^{N-1}I(\br_n;\overline{\bx}_n).
\end{align}


Hence, the ergodic capacity can be obtained as
\begin{align}
C_{\text{OTFS}}\!&= \!C_{\text{OFDM}} \!\!\!\!
&= \!\! \frac{1}{M+M_{\text{cp}}}  \E  \left[\log_2\left(\frac{|\bK_n \bK_n^{\rm H} +\sigma^2\bI_{M{n_{\rm r}}}|}{|\sigma^2\bI_{M{n_{\rm r}}}|}\right)\right].
\end{align}

While in the above analysis we reached the conclusion that OTFS has the same ergodic capacity as OFDM, it is important to note that such analysis assumes perfect knowledge of the (time-varying) channel, an optimal detector, and an infinite code length. In a practical receiver design, however, the sparsity and lower variability of the OTFS channel in the Doppler- delay domain yield great benefits over OFDM, especially for higher Doppler channels. In such scenarios, an OFDM receiver needs to track rapid channel variations in the time-domain. Here, because of the channel variation over each OFDM symbol, keeping track of such variations turns out to be a difficult task. In OTFS, on the other hand, channel variation in time averages out and translates to a much slower variation in the Doppler-delay domain. In addition, it results in a sparse channel which will be easier to estimate. These benefits of OTFS, clearly, enable a simpler channel estimation and equalization design and hence reduce overhead and complexity of the receiver.  

\section{Conclusion}\label{sec:X}
In this work, we conducted a discrete-time analysis for MIMO OFDM-based OTFS modulation. Such analysis led to a concise, vectorized input-output relationship that is applicable to general time-varying channels with arbitrary Dopplers and windowing functions. We provided an accurate characterization of the ergodic capacity which shows that both OFDM and OTFS achieve the same ergodic capacity despite great benefits of the latter in practical receiver design. The analysis developed here provides a strong theoretical foundation for the design of practical detectors/equalizers for OTFS systems.

\bibliographystyle{IEEEtran}
\bibliography{OTFS_Capacity}

\end{document}